\pdfoutput=1

\documentclass[11pt,twoside,a4paper,cmspaper,final,collab]{cms-tdr}
\def\svnVersion{230901}\def\cmsCernNo{2013-037}\def\cmsCernDate{\today}\def\cmsMessage{Published in Physical Review Letters as \href{http://dx.doi.org/10.1103/PhysRevLett.111.211804}{\doi{10.1103/PhysRevLett.111.211804.}}}

\begin{document}\cmsNoteHeader{B2G-13-001}

\hyphenation{had-ron-i-za-tion}
\hyphenation{cal-or-i-me-ter}
\hyphenation{de-vices}

\RCS$Revision: 219222 $
\RCS$HeadURL: svn+ssh://alverson@svn.cern.ch/reps/tdr2/papers/B2G-13-001/trunk/B2G-13-001.tex $
\RCS$Id: B2G-13-001.tex 219222 2013-12-03 19:18:17Z jpilot $
\newlength\cmsFigWidth
\ifthenelse{\boolean{cms@external}}{\setlength\cmsFigWidth{0.95\columnwidth}}{\setlength\cmsFigWidth{0.8\textwidth}}
\ifthenelse{\boolean{cms@external}}{\providecommand{\cmsLeft}{top}}{\providecommand{\cmsLeft}{left}}
\ifthenelse{\boolean{cms@external}}{\providecommand{\cmsRight}{bottom}}{\providecommand{\cmsRight}{right}}
\ifthenelse{\boolean{cms@external}}{\providecommand{\suppMaterial}{the supplemental material [URL will be inserted by publisher]}}{\providecommand{\suppMaterial}{Appendix~\ref{app:suppMat}}}

\renewcommand{\GeVcc}{\GeV}
\renewcommand{\GeVc}{\GeV}
\renewcommand{\TeVcc}{\TeV}
\renewcommand{\TeVc}{\TeV}

\cmsNoteHeader{B2G-13-001} % This is over-written in the CMS environment: useful as preprint no. for export versions
\title{Searches for new physics using the \texorpdfstring{\ttbar}{ttbar} 
invariant mass distribution in \texorpdfstring{$\Pp\Pp$}{pp} collisions at \texorpdfstring{$\sqrt{s}=8\TeV$}{sqrt(s)=8 TeV}}

\author{The CMS Collaboration}

\date{\today}

\abstract{
Searches for anomalous top quark-antiquark production are presented, based
on pp collisions at $\sqrt{s} = 8$\TeV.
The data, corresponding to an integrated luminosity of 19.7\fbinv, were
collected with the CMS detector at the LHC.
The observed \ttbar invariant mass spectrum is found to be compatible 
with the standard model prediction.
Limits on the production cross section times branching fraction probe,
for the first time, a region of parameter space for certain models of new
physics not yet constrained by precision measurements.
}

\hypersetup{%
pdfauthor={CMS Collaboration},%
pdftitle={Searches for new physics using the ttbar invariant mass distribution in pp collisions at sqrt(s)=8 TeV},%
pdfsubject={CMS},%
pdfkeywords={CMS, physics, top production}
}

\maketitle %maketitle comes after all the front information has been supplied

With the discovery of a Higgs boson with a mass around 125\GeVcc~\cite{Aad:2012tfa,Chatrchyan:2012ufa,Chatrchyan:2013lba},
the focus of particle physics has shifted towards
understanding the properties of the new boson, uncovering the
nature of the underlying electroweak symmetry breaking (EWSB)
mechanism, and finding new physics. The standard model (SM) is
believed to be an effective
theory, \ie, a low-energy approximation of a more complete theory
incorporating gravity and explaining the origin of many parameters
that are simply postulated within the SM. Many models beyond the SM (BSM)
have been proposed in order to alleviate the hierarchy problem
of the SM, which stems from the fact that quantum-loop corrections to
the Higgs boson mass diverge quadratically with the highest energy
scale of the model, requiring an enormous degree of fine tuning to
ensure that the Higgs mass remains close to the W-boson mass up to the Planck scale.
Since the largest quantum correction to the Higgs boson mass involves
a top-quark loop, it is natural to suppose that these BSM mechanisms
would involve interactions with the top quark.

Potential solutions to the hierarchy problem include models with extra
spatial dimensions, either flat~\cite{ed} or warped~\cite{rs1,rs2}. In
these models,
gravity is allowed to permeate the multidimensional space, which
results in its apparent weakness from the point of view of
an observer restricted to
3+1 dimensions. This effect lowers the highest energy
scale in the SM from the Planck scale to the \TeVns{} scale, thus eliminating
the hierarchy between the EWSB scale and the highest scale in the
theory.
Such models often contain
Kaluza--Klein excitations of
particles, including gravitons and gluons, both
of which can have enhanced couplings to $\ttbar$
pairs~\cite{Agashe:2006hk}.
Other new gauge bosons have been proposed, referred to generically as
$\PZpr$, that also couple preferentially
to $\ttbar$
pairs~\cite{littlehiggs,zprime_Rosner,ZprimeXS,axigluon,Choudhury:2007ux,pseudohiggs}.
Furthermore, there may be additional spin-zero resonances
that preferentially decay to $\ttbar$ pairs~\cite{pseudohiggs,Frederix:2007gi}.
These various resonances may be observable as enhancements in the \ttbar 
invariant mass spectrum.

Discrepancies
have been observed in the forward-backward asymmetry
of top quark production at the
Tevatron~\cite{Lannon:2012fp}.
Assuming this anomaly is due to new physics above the \TeVns{} scale, an
enhancement of the $\ttbar$ rate at high invariant mass could be visible at
the Large Hadron Collider (LHC)~\cite{top_afb_implications2,top_afb_implications1}.

In this Letter, a search for anomalous production of $\ttbar$ events
is presented, from data corresponding to an integrated luminosity of
$19.7\fbinv$ of pp collisions at  $\sqrt{s}=8\TeV$, recorded with
the Compact Muon Solenoid (CMS) detector~\cite{cmsdetector} at the LHC.
These results
represent a significant improvement over the previous
searches~\cite{Deliot:2013gla,cms-allhad,atlas-ljets,atlas-ljets-boosted,cms-ljets,cms-dilep},
due primarily
to the large
increase in the high-$x$ parton luminosity from the higher
LHC energy in 2012, but also because of the
increased size of the data sample and the combination
of several statistically-independent channels.
Specific comparisons are made to the resonant production of
Randall--Sundrum Kaluza--Klein
(RS KK) gluons~\cite{Agashe:2006hk}, of a $\PZpr$ boson in the
topcolor model~\cite{ZprimeXS}, and of a scalar Higgs-like boson
produced via
gluon fusion through its couplings to the top quarks.
In addition, enhancements of the $\ttbar$ invariant mass
($M_{\ttbar}$) spectrum are constrained for $M_{\ttbar} > 1\TeVcc$.
These results probe,
for
the first time, a region of parameter space of models with warped
extra dimensions not yet constrained by precision measurements~\cite{Davoudiasl:2009cd}.

Since the top quark decays primarily to a $\PW$ boson and a bottom
($\cPqb$) quark, top pair production signatures are classified based
on whether the $\PW$ bosons decay to leptons or quarks.
This measurement combines analyses utilizing the final states
where one or both $\PW$ bosons from $\ttbar$ events decay to quarks
(``semi-leptonic'' and ``all-hadronic'' events, respectively).
The events are classified into two categories
based on the expected kinematics of the top-quark decay products.
In the first category, the \ttbar
pair is produced near the kinematic threshold, resulting in
a topology where each parton is matched to a single
jet (``resolved topology'').
In the second category, each top quark is
produced with a high Lorentz boost (${>}2$), resulting in
collimated decay products that may be clustered into a single
jet (``boosted topology'').
The transition between the resolved and boosted topologies occurs
around $M_{\ttbar} = 1$\TeVcc.
Both the resolved and boosted topologies are used
to analyze the semi-leptonic events. However, all-hadronic events
are analyzed only in the boosted topology, which is combined with the
semi-leptonic boosted events to perform the search.  
As the
all-hadronic analysis is dominated by multijet events, jet
substructure criteria are imposed to further enhance sensitivity.
The analysis techniques are similar to those explored in earlier
analyses of $\Pp\Pp$ collision data~\cite{cms-ljets,cms-allhad}.

The CMS detector, a general-purpose apparatus operating at the CERN
LHC, is described in detail elsewhere~\cite{cmsdetector}.
The central feature of the CMS apparatus is a superconducting solenoid of 6\unit{m} internal diameter,
providing a magnetic field of 3.8\unit{T}. Within the superconducting solenoid volume are a silicon pixel and strip tracker,
a lead tungstate crystal electromagnetic calorimeter, and a brass/scintillator hadron calorimeter. Muons are
measured in gas-ionization detectors embedded in the steel magnetic flux return yoke
outside the solenoid.

At the CMS experiment,
the polar angle $\theta$ is measured from the beam direction and
the azimuthal angle $\phi$ is measured perpendicular to the beam direction.
The rapidity $y$ is approximated by the pseudorapidity $\eta$,
defined as $\eta = -\ln[\tan(\theta/2)]$.
The transverse momentum perpendicular to the beamline is denoted
as $\pt$.

The CMS experiment uses a particle-flow-based event reconstruction~\cite{pfalgo},
which aggregates information from all subdetectors, including charged-particle tracks from the
tracking system and deposited energy from the electromagnetic and hadron calorimeters. Given this
information, all detected particles in the event are reconstructed as electrons, muons,
photons, neutral hadrons, or charged hadrons.
For this paper, electrons and photons are required to have
$\abs{\eta} < 2.5$, and muons, $\abs{\eta} < 2.1$.
The leading hard-scattering vertex of the event is defined as the
vertex whose tracks have the largest squared-sum of transverse
momentum.
Charged hadrons associated with other vertices
are removed from further consideration.
The remaining candidates are
clustered into jets using the \textsc{FastJet} 3.0 software
package~\cite{FastJet}.
The semi-leptonic analyses use the anti-\kt
jet-clustering algorithm~\cite{Cacciari:2008gp} with a size parameter
of 0.5 (AK5 jets),
while the all-hadronic analysis uses the Cambridge-Aachen (CA)
jet-clustering algorithm~\cite{CACluster1,CACluster2} with a size
parameter of 0.8 (CA8 jets) to take advantage of the capability of the
CA algorithm to distinguish jet substructure.
Jets
are required to
satisfy $\abs{\eta} < 2.4$. Jets are identified as $\cPqb$-quark jets
if they satisfy the combined secondary vertex
algorithm defined in Ref.~\cite{Chatrchyan:2012jua}.

The data for the semi-leptonic resolved category were collected with
triggers requiring a single isolated muon or electron
with a $\pt$ threshold of $17$ or $25$\GeVc, respectively, in combination
with three jets with $\pt > 30$\GeVc.
Off\-line, we select events containing exactly one isolated muon
(with $\pt^\mu > 26$\GeVc), or electron (with
$\pt^\Pe > 30$\GeVc), and
at least four jets with
$\pt>70$, 50, 30, 30\GeVc, respectively.
The non-$\PW$ multijet background (NWMJ) is suppressed further
by requiring the transverse missing momentum $\MET$, the
modulus of the vector sum of all measured particle $\pt$, to be larger
than 20\GeVc.

For the semi-leptonic boosted category, data were recorded with triggers requiring
one muon ($\pt^\Pgm > 40$\GeVc), or one electron ($\pt^\Pe > 35$\GeVc) in
conjunction with two
jets ($\pt >100$, $25$\GeVc).
Since the top-quark decay products can be collinear in this
regime, no isolation requirements on
the leptons are imposed in either the trigger or offline selections.
Offline, we select events containing exactly one muon with
$\pt^\Pgm > 45$\GeVc, or exactly
one electron with $\pt^\Pe > 35$\GeVc and at least two jets with $\pt >
150$, 50\GeVc, respectively.
To reduce the contamination of NWMJ processes, we
follow the techniques of Ref.~\cite{cms-ljets}, placing requirements
on the angle and relative momentum between the lepton and the nearest
jet, and also requiring $\MET > 50\GeVc$
and $\MET + \pt^{\Pe,\Pgm} > 150$\GeVc.

The events satisfying the two semi-leptonic selections are separated
into
categories determined by the lepton flavor (electron or muon) and the
number of $\cPqb$-tagged jets $N_{\text{b-tag}}$ (1 or $\geq$2
$\cPqb$-tagged jets for the resolved analysis, and
0 or $\geq$1 $\cPqb$-tagged jets for the boosted analysis).
The purpose of this classification is to separate the sample into
regions dominated by different background processes. The events in the
categories with fewer $\cPqb$-tagged jets have a higher fraction of
$\PW+$jets events, whereas those with
more
$\cPqb$-tagged jets have a higher fraction of $\ttbar$ events. This
characterization
is used to constrain the various background components
by imposing self-consistency among the channels.
The reconstruction of semi-leptonic $\ttbar$
candidates relies on a $\chi^2$ variable built by enforcing
kinematic consistency (within uncertainties) with the $\ttbar$
hypothesis, imposing constraints on the reconstructed $\PW$ and
top candidates.
In the semi-leptonic boosted regime we follow the techniques of
Ref.~\cite{cms-ljets}, and allow candidates with more than one parton
merged into a single jet.

For the boosted all-hadronic analysis, data were recorded with a trigger requiring the
scalar sum of the
transverse momenta of reconstructed AK5 jets to be greater
than 750$\GeVc$.  In the offline
analysis selection, we require
two CA8 jets, each with $\pt > 400\GeVc$.%, chosen to be in the plateau

The reconstruction of the boosted all-hadronic analysis 
relies on ``top-tagging'' techniques similar to those used in the 
previous analysis~\cite{cms-allhad}.
This algorithm~\cite{TopTagAlgo}, aiming to identify the top quark decay products
within CA8 jets, reverses the jet-clustering sequence by iteratively
separating the jet into subjets 
until three or four subjets with sufficient $\pt$ are found.
The algorithm is validated on a sample of \ttbar events selected by requiring
one muon and additional jets
\ifthenelse{\boolean{cms@external}}{~\cite{suppMaterial}.}
{as shown in Fig.~\ref{fig:mass} in the Appendix.}
The reconstructed single-jet top quark mass is found to be consistent with the 
expectation from simulated events, as is the
di-subjet $\PW$ mass, obtained from the minimum mass pairing of the 
leading three subjets.
The two selected top-tagged jets are then required to be 
back-to-back, with  $\abs{\Delta \phi} > \pi / 2$ and $\abs{\Delta y} < 1.0$,
to suppress non-top multijet (NTMJ) backgrounds.

SM top-quark production is modeled with the next-to-leading-order (NLO) generator
\POWHEG (v1.0)~\cite{powheg}, interfaced with \PYTHIA6 (v6.2.24)~\cite{pythia} for
parton showering with tune {Z2$^*$}~\cite{Chatrchyan:2011id}.
\MADGRAPH (v5.1.1)~\cite{madgraph}
interfaced to \PYTHIA6
is used for simulating $\PW$ and $\cPZ$ boson production in association with jets.
Diboson processes
are generated with
\PYTHIA6 to compute both the matrix element and showering.

The \MADGRAPH --\PYTHIA6 combination is also used to generate
signal Monte Carlo (MC) simulation events for limit setting, including
high-mass SM-like $\PZpr$ resonances with $\Gamma_{\PZpr}/M_{\PZpr} = 1$\% and
$\Gamma_{\PZpr}/M_{\PZpr} = 10$\%, where $\Gamma_{\PZpr}$ is the
width of the resonance, and $M_{\PZpr}$ is the mass.
This relative width can be compared to the detector resolution of about 10\% for a
$\ttbar$ resonance mass. Hence, limits set for the $\PZpr$ with a
width of 1\% would apply to a larger class of models in which the resonance width is
below the experimental resolution. The \MADGRAPH --\PYTHIA6
combination is also used to
generate a simplified model of a spin-zero resonance produced via
gluon fusion through its couplings to top quarks, with the SM
interference effects neglected in the model. A Kaluza--Klein excitation
of a gluon with a width
of approximately 15-20\%~\cite{Agashe:2006hk} is generated with \PYTHIA8
(v1.5.3)~\cite{pythia8}.

The leading-order (LO) CTEQ6L parton distribution functions~\cite{Nadolsky:2008zw} are used,
except for the generation of the \POWHEG samples, which use the
NLO CT10 parton distribution function~\cite{CT10}.
All MC samples include additional collisions per beam crossing and 
are reweighted to match the data taking conditions as well as the 
identification and trigger efficiencies measured in control samples.

For the semi-leptonic resolved analysis, an aggregate
background estimate is taken for all SM components together directly
from the data, with the
SM $\ttbar$ component the dominant one. The
number of signal events is extracted from a 
binned maximum likelihood fit to the $M_{\ttbar}$
distribution, assuming a
smoothly-falling probability density function (pdf) for the SM
backgrounds and a
parameterization of the signal pdf based on a Breit--Wigner shape.
Only events with $M_{\ttbar} > 550$\GeVcc are considered; below this
value the SM backgrounds are not described by a smoothly-falling pdf.

The $M_{\ttbar}$ distributions of the semi-leptonic
and all-hadronic boosted
topologies are fitted together in a single joint likelihood maximization,
imposing consistency of the various background and signal components
across the two channels.
The initial estimates for the
SM \ttbar, single-top-quark, $\PW+$jets, and diboson production
are based on simulation after applying data-MC corrections based on
control samples in data.
The boosted all-hadronic analysis has one additional background component,
the NTMJ background,
which is estimated using the probability to misidentify a
light-parton jet as a top-quark jet measured in data.
This probability is derived as a function of the jet $\pt$ in a sample 
enriched in light-quark jets, kinematically similar to the signal region.
It is then used to weight events
in the signal region. Furthermore, the efficiency for identifying true
top-quark jets is corrected in the signal MC simulations using measurements in a 
signal-depleted sideband region containing events with one isolated muon and additional jets.
It is found that the efficiencies in data and MC simulations agree, having a ratio of $93\pm5\%$.
The methods described above were validated using simulated samples and it was 
verified that signal contamination was minimal in the signal-depleted regions.

In the likelihood maximization, systematic
uncertainties are treated as nuisance parameters. Those that are
common among the channels are treated as 100\%
correlated, while those that are
channel-specific are treated as uncorrelated.
The normalizations of the backgrounds are
allowed to vary within %conservative
log-normal constraints in the maximization of the joint likelihood.
The shapes of the backgrounds are also allowed to vary within their
uncertainties. The shapes and normalizations also account for
systematic variations due to efficiency and misidentification
rates. The constraints used in the joint likelihood
maximization are listed in Table~\ref{tab:sys}.

\begin{table}
  \centering
  \topcaption{\label{tab:sys}Constraints used in the likelihood maximization.
  The $M_{\ttbar}$ distributions of the boosted channels are combined into a single joint likelihood, imposing consistency of the various background and signal components.}
    \begin{scotch}{lccc}
                       & Resolved     & Boosted         & Boosted \\
                       & Semi- & Semi-   &  All- \\
                       & Leptonic & Leptonic & Hadronic \\
\hline
\multicolumn{4}{c}{ Constraints on normalization} \\
\hline
Luminosity~\cite{LumiUncertainty} & 2.6\% & 2.6\%  & 2.6\%  \\
Pileup & 6\%   & 6\%    & 6\%  \\
$\ttbar$ ~\cite{PhysRevD.84.092004} & -- & 15\% & 15\%  \\
Parton distribution & \multirow{2}{*}{$ 1 \sigma$ } & \multirow{2}{*}{${1}\sigma$} & \multirow{2}{*}{${1}\sigma$} \\
\quad functions~\cite{Pumplin:2001ct} &  &  & \\
Single top    & -- &  50\% & -- \\
$\PW$+light-flavor jets & -- & 50\%  & -- \\
$\PW$+heavy-flavor jets & -- & 100\% & -- \\
$\Z$+ jets & -- & 100\% & -- \\
Lepton selection & 0.5-3.0\% & 0.5-3.0\% & -- \\
Top-tagging efficiency & -- & -- &  9\% \\

\hline
\multicolumn{4}{c}{ Constraints on shape } \\
\hline
$\ttbar$ renormalization, & \multirow{3}{*}{--} & \multicolumn{2}{c}{\multirow{3}{*}{variation by $\times 2$ and $\times 0.5$}} \\
\quad factorization and & & &  \\
\quad matching scales & & & \\
Jet energy scale & 1--6\% & 1--6\% & 1--6\%  \\
Jet energy resolution & 8--10\% & 8--10\% & 8--10\% \\
$\cPqb$-tagging efficiency & 2--8\% & 2--8\% & -- \\
$\cPqb$-tagging  mis-ID  & -- & 20\% & --\\
Top-tagging mis-ID & -- & -- & 5--20\% \\
Signal and & \multirow{2}{*}{$1 \sigma$} & \multirow{2}{*}{--} & \multirow{2}{*}{--} \\
\quad background pdf & & & \\
\end{scotch}

\end{table}

\begin{figure}[htbp]
\centering
\includegraphics[width=0.48\columnwidth]{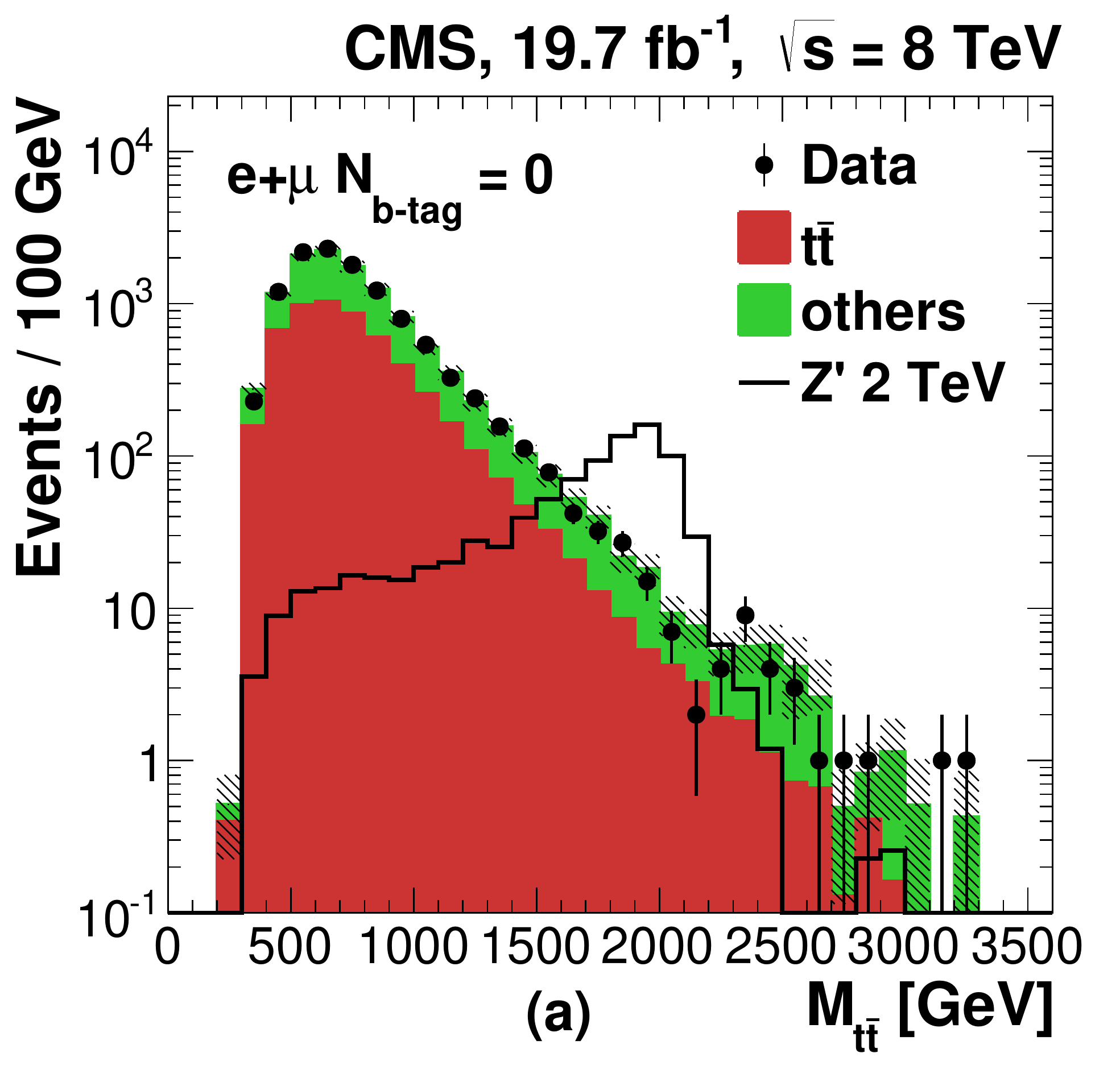}
\includegraphics[width=0.48\columnwidth]{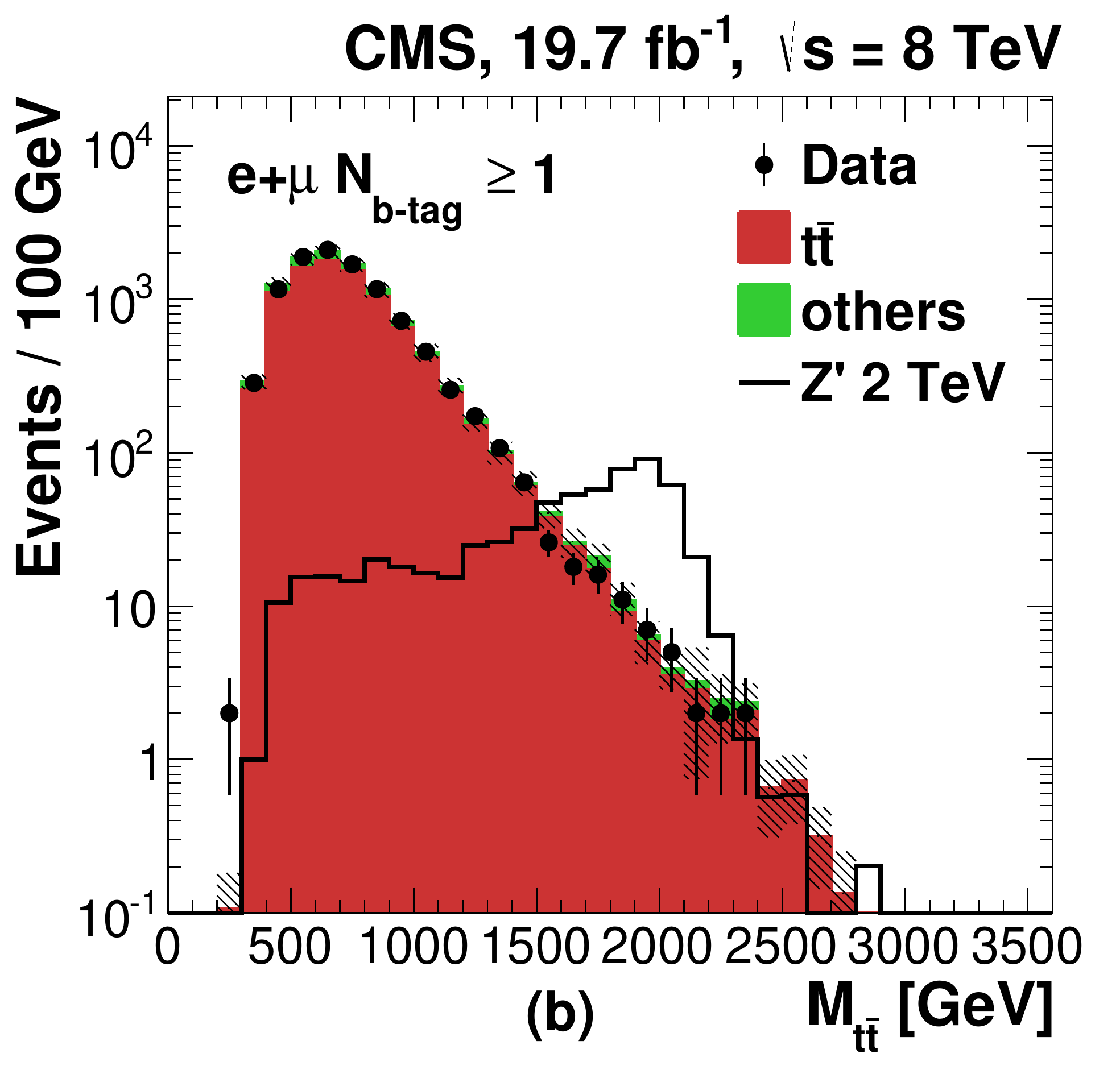}
\includegraphics[width=0.48\columnwidth]{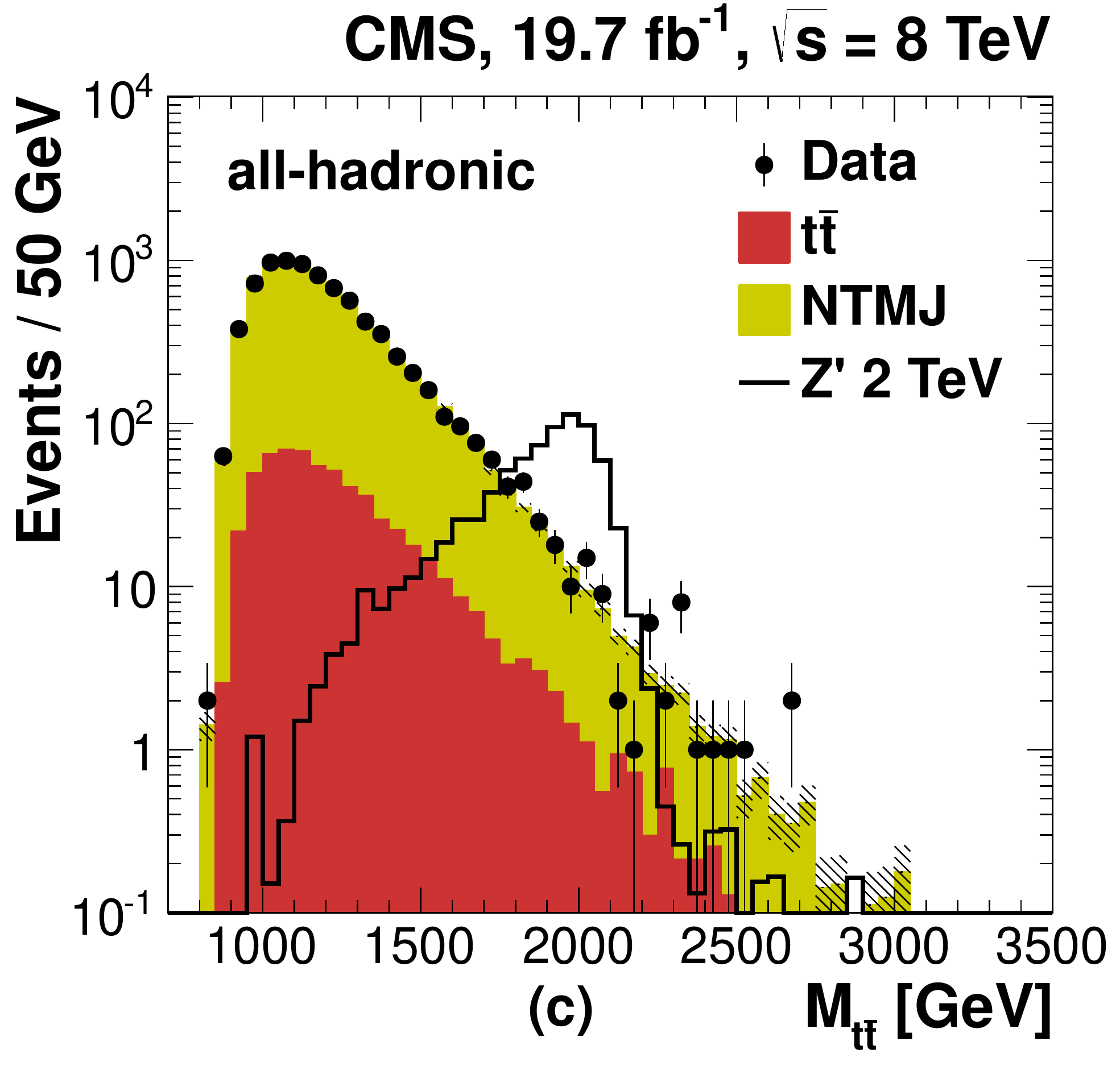}
\includegraphics[width=0.48\columnwidth]{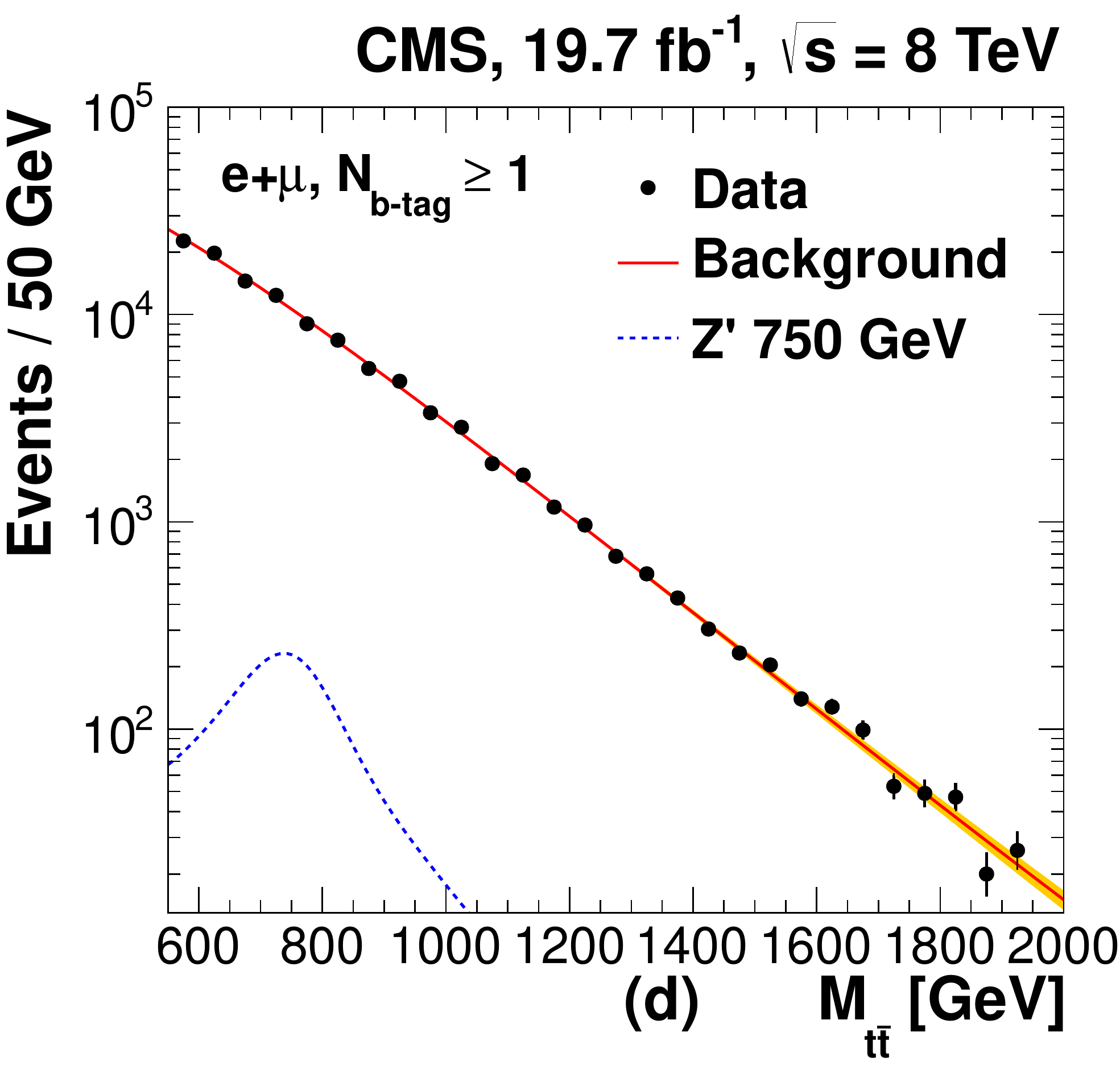}
\caption{
Comparison between data and SM prediction for reconstructed
$M_{\ttbar}$ distributions for the boosted semi-leptonic analysis with 0
b-tagged jets (a) and ${\ge}1$ b-tagged jets (b), as well as for the
all-hadronic analysis (c). For the semi-leptonic analyses,
``others'' refers to all non-top backgrounds, while for the
all-hadronic analysis, ``NTMJ''
refers to the ``non-top multijet'' background.
The shaded band corresponds to
the SM background uncertainty.
The likelihood fit projection on data for the semi-leptonic resolved
analysis is shown in (d).
A cross section of $1.0\unit{pb}$ is used for the normalization of the $\PZpr$ samples.
\label{fig:mttbar}}
\end{figure}

The event yields from the various background components and data are
shown in Table~\ref{tab:yields}.
The yields of the simulated samples are quoted
after the likelihood maximization procedure, and the individual background
uncertainties include
only the uncertainty in the individual normalization. The total SM
contribution includes all
uncertainties, including the correlations not quoted in the individual
components.
Figure~\ref{fig:mttbar} shows the $M_{\ttbar}$ distributions for all
channels along with the expectation from a $\PZpr$ signal.

\begin{table}
  \centering
  \topcaption{\label{tab:yields} Number of expected and
       observed events in the
       boosted analyses. 
}
      \begin{scotch}{lccc}
        Sample & Semi-Leptonic & Semi-Leptonic  & All-Hadronic \\
               & $N_{\text{b-tag}}=0$ & $N_{\text{b-tag}} \ge 1$ & $M_{\ttbar} \ge 1\TeVcc$ \\
       \hline
        $\ttbar$ &   5440 $\pm$  520  &  9090 $\pm$ 870  & 510  $\pm$ 90  \\
        NTMJ           &    --              &   --             & 6600 $\pm$ 200 \\
        Others         &   5880 $\pm$  820  &  1070 $\pm$ 380  & --             \\
        Total SM       &  11320 $\pm$ 1300  & 10160 $\pm$ 1300 & 7110 $\pm$ 410 \\
        Data           &  10305             & 10159            & 6887 \\
      \end{scotch}
\end{table}

In all cases, the data are well-described by the SM-only background
hypothesis.
The absence of a signal in the $M_{\ttbar}$ distribution is
quantified by deriving Bayesian upper limits on the signal cross
section times branching fraction at 95\% confidence level (CL), using
pseudo-experiments.
The resolved semi-leptonic analysis has some overlapping phase
space with the boosted semi-leptonic analysis, and there is a transition point
(${\sim}1$\TeVcc) where the expected sensitivities of the boosted and resolved
analyses are equal, above which the boosted
analysis result is quoted,
and below which the resolved
analysis result is quoted.

\begin{figure}[htbp]
\centering
\includegraphics[width=\cmsFigWidth]{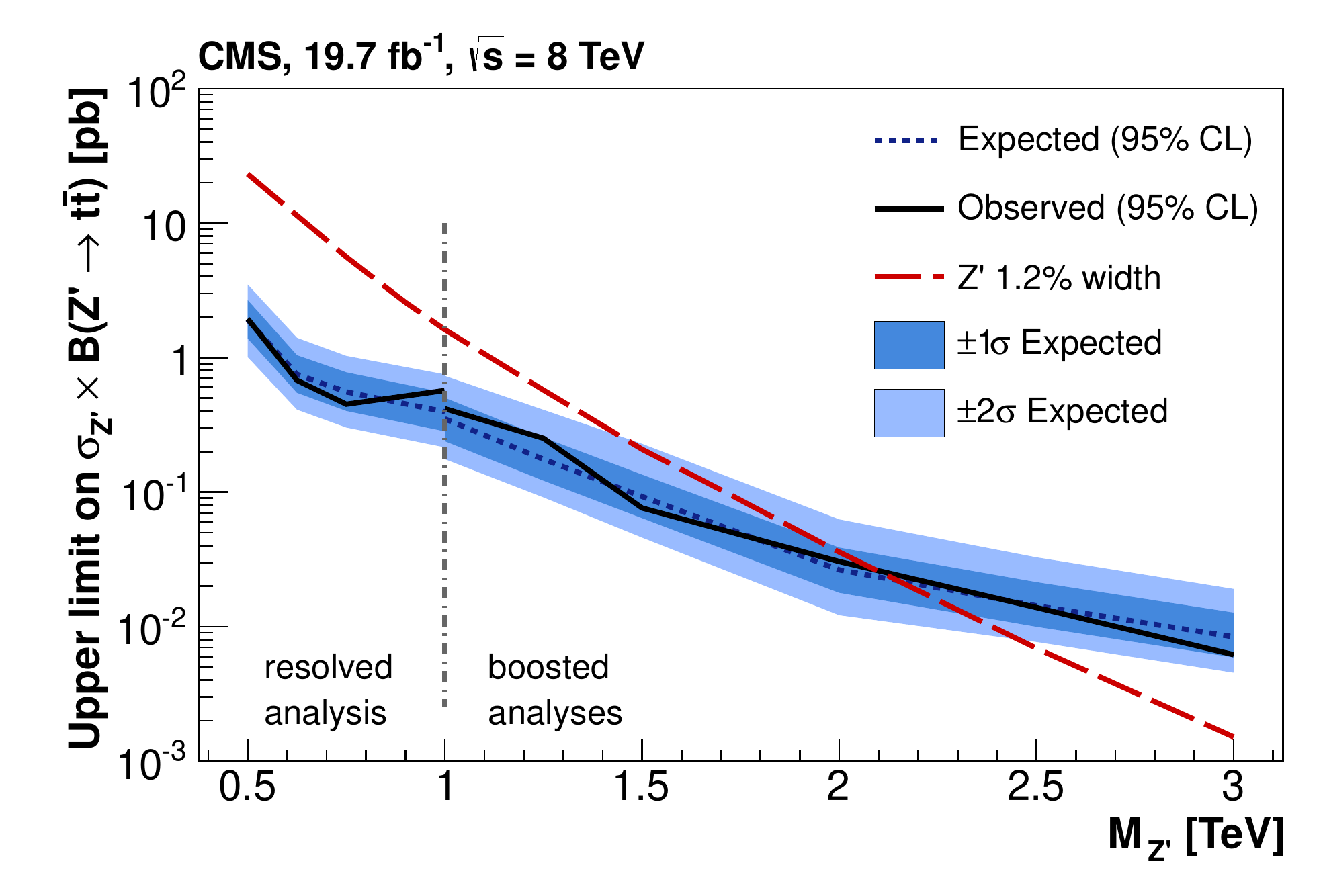}
 \caption{The 95\% CL upper limits on the production
 cross section times branching fraction as a function of $M_{\ttbar}$ for
$\PZpr$ resonances  with $\Gamma_{\PZpr}/M_{\PZpr}=1.2\%$
compared to
predictions from Ref.~\cite{ZprimeXS} multiplied by 1.3 to account
for higher-order
effects~\cite{Gao:2010bb}. 
}
\label{fig:limits}
\end{figure}

Figure~\ref{fig:limits} shows the expected and
observed limits
for a narrow resonance,
as a function of the invariant mass of the resonance.
The specific example shown in Fig.~\ref{fig:limits} and given by the
dashed line refers to a topcolor $\PZpr$ with $\Gamma_{\PZpr}/M_{\PZpr}=1.2$\%
based on predictions from Ref.~\cite{ZprimeXS}.
The cross-section limits for this case
are obtained from the MC models with
 $\Gamma_{\PZpr}/M_{\PZpr}=1.0$\%, scaled by the ratio of
 theoretical cross sections. This scaling is done to compare to theoretical
 results and previous measurements.
As the cross section calculation is available for this model at LO
only, the predictions are
multiplied by a factor of 1.3 to account for higher-order
effects~\cite{Gao:2010bb}. The vertical dash-dotted line indicates the
transition between the resolved and boosted analyses.
Table~\ref{tab:limits} shows additional model-specific limits.
The combination of the semi-leptonic and all-hadronic boosted analyses
improves the expected cross section limits at 2\TeVcc by ${\sim}25$\%.
Compared to the
results of previous
analyses~\cite{cms-allhad,atlas-ljets,atlas-ljets-boosted,cms-ljets}
for specific
models~\cite{ZprimeXS, Agashe:2006hk},
the lower limits on the masses of these resonances have been improved
by several hundred $\GeVns$.
For the semi-leptonic resolved analysis,
assuming a spin-zero resonance with narrow width, produced via gluon
fusion with no interference with the
SM background,
the cross section limits are
0.8\unit{pb} and 0.3\unit{pb} for a spin-zero resonance of mass 500\GeVcc and 750\GeVcc,
respectively.
These are the first limits at CMS for heavy Higgs-like particles decaying
into $\ttbar$.

\begin{table}
\centering
\topcaption{\label{tab:limits} 95\% CL lower limits on the masses of new
  particles in specific models.}
    \begin{scotch}{lccc}
       Model & Observed Limit & Expected Limit \\
      \hline
$\PZpr$, $\Gamma_{\PZpr}/M_{\PZpr}=$ 1.2\% & 2.1\TeVcc & 2.1\TeVcc \\
$\PZpr$, $\Gamma_{\PZpr}/M_{\PZpr}=$ 10\% & 2.7\TeVcc & 2.6\TeVcc \\
RS KK gluon & 2.5\TeVcc & 2.4\TeVcc \\
    \end{scotch}

\end{table}

In addition to investigating possible resonant structures in the
$M_{\ttbar}$ spectrum,
the presence of new physics that causes a
non-resonant enhancement of the $M_{\ttbar}$ spectrum is
also tested.
The boosted all-hadronic analysis is used to set limits on
such new production for events with $M_{\ttbar} > 1$\TeVcc, since the
NTMJ background can be predicted entirely from data.
The limit is expressed as a
ratio of the total SM + BSM $\ttbar$ cross section to the SM-only
cross section ($\mathcal{S}$, as defined in Ref.~\cite{cms-allhad}).
The efficiency to select SM \ttbar events
with $M_{\ttbar} > 1$\TeVcc
is
$(3.4 \pm 1.7) \times 10^{-4}$.  We find
$\mathcal{S} < 1.2 $ at the 95\% CL,
with a credible interval
of 1.1--2.0 at
68\% CL, a factor of two improvement over the previously published
limits~\cite{cms-allhad}.

In summary, we have performed
searches for anomalous $\ttbar$ production using events in the
semi-leptonic and all-hadronic topologies. In addition to new limits
on nonresonant enhancements to top-quark
production, limits are set on the production cross section times
branching fraction for several resonance hypotheses, for resonances in
the mass range 0.5--3.0\TeVcc.

We congratulate our colleagues in the CERN accelerator departments for the excellent performance of the LHC and thank the technical and administrative staffs at CERN and at other CMS institutes for their contributions to the success of the CMS effort. In addition, we gratefully acknowledge the computing centers and personnel of the Worldwide LHC Computing Grid for delivering so effectively the computing infrastructure essential to our analyses. Finally, we acknowledge the enduring support for the construction and operation of the LHC and the CMS detector provided by the following funding agencies: BMWF and FWF (Austria); FNRS and FWO (Belgium); CNPq, CAPES, FAPERJ, and FAPESP (Brazil); MES (Bulgaria); CERN; CAS, MoST, and NSFC (China); COLCIENCIAS (Colombia); MSES (Croatia); RPF (Cyprus); MoER, SF0690030s09 and ERDF (Estonia); Academy of Finland, MEC, and HIP (Finland); CEA and CNRS/IN2P3 (France); BMBF, DFG, and HGF (Germany); GSRT (Greece); OTKA and NKTH (Hungary); DAE and DST (India); IPM (Iran); SFI (Ireland); INFN (Italy); NRF and WCU (Republic of Korea); LAS (Lithuania); CINVESTAV, CONACYT, SEP, and UASLP-FAI (Mexico); MBIE (New Zealand); PAEC (Pakistan); MSHE and NSC (Poland); FCT (Portugal); JINR (Dubna); MON, RosAtom, RAS and RFBR (Russia); MESTD (Serbia); SEIDI and CPAN (Spain); Swiss Funding Agencies (Switzerland); NSC (Taipei); ThEPCenter, IPST, STAR and NSTDA (Thailand); TUBITAK and TAEK (Turkey); NASU (Ukraine); STFC (United Kingdom); DOE and NSF (USA).

\bibliography{auto_generated}   % will be created by the tdr script.

\providecommand{\href}[2]{#2}\begingroup\raggedright\begin{thebibliography}{10}%
\makeatletter
\providecommand{\hrefCMSnoop }[0]{\@secondoftwo}%
\makeatother
\providecommand{\doi}{\texttt{doi:}\begingroup \urlstyle{tt}\Url}

\bibitem{Aad:2012tfa}
\hrefCMSnoop {} {{ ATLAS} Collaboration, ``{Observation of a new particle in
  the search for the Standard Model Higgs boson with the ATLAS detector at the
  LHC}'',} \textit{ Phys. Lett. B} \textbf{ 716} (2012) 1,
  \href{http://dx.doi.org/10.1016/j.physletb.2012.08.020}{\doi{10.1016/j.physletb.2012.08.020}},
\href{http://www.arXiv.org/abs/1207.7214}{\texttt{ arXiv:1207.7214}}.
%%CITATION = ARXIV:1207.7214;%%.

\bibitem{Chatrchyan:2012ufa}
\hrefCMSnoop {} {{ CMS} Collaboration, ``{Observation of a new boson at a mass
  of 125 GeV with the CMS experiment at the LHC}'',} \textit{ Phys. Lett. B}
  \textbf{ 716} (2012) 30,
  \href{http://dx.doi.org/10.1016/j.physletb.2012.08.021}{\doi{10.1016/j.physletb.2012.08.021}},
\href{http://www.arXiv.org/abs/1207.7235}{\texttt{ arXiv:1207.7235}}.
%%CITATION = ARXIV:1207.7235;%%.

\bibitem{Chatrchyan:2013lba}
\hrefCMSnoop {} {{ CMS} Collaboration, ``{Observation of a new boson with mass
  near 125 GeV in pp collisions at $\sqrt{s} = $ 7 and 8 TeV}'',} \textit{
  JHEP} \textbf{ 06} (2013) 081,
  \href{http://dx.doi.org/10.1007/JHEP06(2013)081}{\doi{10.1007/JHEP06(2013)081}},
\href{http://www.arXiv.org/abs/1303.4571}{\texttt{ arXiv:1303.4571}}.
%%CITATION = ARXIV:1303.4571;%%.

\bibitem{ed}
\hrefCMSnoop {} {N.~Arkani-Hamed, S.~Dimopoulos, and G.~R. Dvali, ``{The
  hierarchy problem and new dimensions at a millimeter}'',} \textit{ Phys.
  Lett. B} \textbf{ 429} (1998) 263,
  \href{http://dx.doi.org/10.1016/S0370-2693(98)00466-3}{\doi{10.1016/S0370-2693(98)00466-3}},
\href{http://www.arXiv.org/abs/hep-ph/9803315}{\texttt{ arXiv:hep-ph/9803315}}.
%%CITATION = HEP-PH/9803315;%%.

\bibitem{rs1}
\hrefCMSnoop {} {L.~Randall and R.~Sundrum, ``{A large mass hierarchy from a
  small extra dimension}'',} \textit{ Phys. Rev. Lett.} \textbf{ 83} (1999)
  3370,
  \href{http://dx.doi.org/10.1103/PhysRevLett.83.3370}{\doi{10.1103/PhysRevLett.83.3370}},
\href{http://www.arXiv.org/abs/hep-ph/9905221}{\texttt{ arXiv:hep-ph/9905221}}.
%%CITATION = HEP-PH/9905221;%%.

\bibitem{rs2}
\hrefCMSnoop {} {L.~Randall and R.~Sundrum, ``{An alternative to
  compactification}'',} \textit{ Phys. Rev. Lett.} \textbf{ 83} (1999) 4690,
  \href{http://dx.doi.org/10.1103/PhysRevLett.83.4690}{\doi{10.1103/PhysRevLett.83.4690}},
\href{http://www.arXiv.org/abs/hep-th/9906064}{\texttt{ arXiv:hep-th/9906064}}.
%%CITATION = HEP-TH/9906064;%%.

\bibitem{Agashe:2006hk}
K.~Agashe\hrefCMSnoop {} { {et~al.}, ``{CERN LHC} signals from warped extra
  dimensions'',} \textit{ Phys. Rev. D} \textbf{ 77} (2008) 015003,
  \href{http://dx.doi.org/10.1103/PhysRevD.77.015003}{\doi{10.1103/PhysRevD.77.015003}},
\href{http://www.arXiv.org/abs/hep-ph/0612015}{\texttt{ arXiv:hep-ph/0612015}}.
%%CITATION = HEP-PH/0612015;%%.

\bibitem{littlehiggs}
\hrefCMSnoop {} {N.~Arkani-Hamed, A.~G. Cohen, and H.~Georgi, ``{Electroweak
  symmetry breaking from dimensional deconstruction}'',} \textit{ Phys. Lett.
  B} \textbf{ 513} (2001) 232,
  \href{http://dx.doi.org/10.1016/S0370-2693(01)00741-9}{\doi{10.1016/S0370-2693(01)00741-9}},
\href{http://www.arXiv.org/abs/hep-ph/0105239}{\texttt{ arXiv:hep-ph/0105239}}.
%%CITATION = HEP-PH/0105239;%%.

\bibitem{zprime_Rosner}
\hrefCMSnoop {} {J.~L. Rosner, ``{Prominent decay modes of a leptophobic
  $Z^\prime$}'',} \textit{ Phys. Lett. B} \textbf{ 387} (1996) 113,
  \href{http://dx.doi.org/10.1016/0370-2693(96)01022-2}{\doi{10.1016/0370-2693(96)01022-2}},
\href{http://www.arXiv.org/abs/hep-ph/9607207}{\texttt{ arXiv:hep-ph/9607207}}.
%%CITATION = HEP-PH/9607207;%%.

\bibitem{ZprimeXS}
\hrefCMSnoop {} {R.~M. Harris and S.~Jain, ``{Cross sections for leptophobic
  topcolor Z$^\prime$ decaying to top-antitop}'',} \textit{ Eur. Phys. J. C}
  \textbf{ 72} (2012) 2072,
  \href{http://dx.doi.org/10.1140/epjc/s10052-012-2072-4}{\doi{10.1140/epjc/s10052-012-2072-4}},
\href{http://www.arXiv.org/abs/1112.4928}{\texttt{ arXiv:1112.4928}}.
%%CITATION = ARXIV:1112.4928;%%.

\bibitem{axigluon}
\hrefCMSnoop {} {P.~H. Frampton and S.~L. Glashow, ``Chiral color: An
  alternative to the standard model'',} \textit{ Phys. Lett. B} \textbf{ 190}
  (1987) 157,
  \href{http://dx.doi.org/10.1016/0370-2693(87)90859-8}{\doi{10.1016/0370-2693(87)90859-8}}.

\bibitem{Choudhury:2007ux}
\hrefCMSnoop {} {D.~Choudhury, R.~M. Godbole, R.~K. Singh, and K.~Wagh, ``{Top
  production at the Tevatron/LHC and nonstandard, strongly interacting spin one
  particles}'',} \textit{ Phys. Lett. B} \textbf{ 657} (2007) 69,
  \href{http://dx.doi.org/10.1016/j.physletb.2007.09.057}{\doi{10.1016/j.physletb.2007.09.057}},
\href{http://www.arXiv.org/abs/0705.1499}{\texttt{ arXiv:0705.1499}}.
%%CITATION = ARXIV:0705.1499;%%.

\bibitem{pseudohiggs}
\hrefCMSnoop {} {D.~Dicus, A.~Stange, and S.~Willenbrock, ``{Higgs decay to top
  quarks at hadron colliders}'',} \textit{ Phys. Lett. B} \textbf{ 333} (1994)
  126,
  \href{http://dx.doi.org/10.1016/0370-2693(94)91017-0}{\doi{10.1016/0370-2693(94)91017-0}},
  \href{http://www.arXiv.org/abs/hep-ph/9404359}{\texttt{
  arXiv:hep-ph/9404359}}.

\bibitem{Frederix:2007gi}
\hrefCMSnoop {} {R.~Frederix and F.~Maltoni, ``{Top pair invariant mass
  distribution: A window on new physics}'',} \textit{ JHEP} \textbf{ 01} (2009)
  047,
  \href{http://dx.doi.org/10.1088/1126-6708/2009/01/047}{\doi{10.1088/1126-6708/2009/01/047}},
\href{http://www.arXiv.org/abs/0712.2355}{\texttt{ arXiv:0712.2355}}.
%%CITATION = ARXIV:0712.2355;%%.

\bibitem{Lannon:2012fp}
\hrefCMSnoop {} {K.~Lannon, F.~Margaroli, and C.~Neu, ``{Measurements of the
  production, decay and properties of the top quark: A review}'',} \textit{
  Eur. Phys. J. C} \textbf{ 72} (2012) 2120,
  \href{http://dx.doi.org/10.1140/epjc/s10052-012-2120-0}{\doi{10.1140/epjc/s10052-012-2120-0}},
\href{http://www.arXiv.org/abs/1201.5873}{\texttt{ arXiv:1201.5873}}.
%%CITATION = ARXIV:1201.5873;%%.

\bibitem{top_afb_implications2}
\hrefCMSnoop {} {J.~A. Aguilar-Saavedra and M.~Perez-Victoria, ``{Probing the
  Tevatron $\ttbar$ asymmetry at LHC}'',} \textit{ JHEP} \textbf{ 05} (2011)
  034,
  \href{http://dx.doi.org/10.1007/JHEP05(2011)034}{\doi{10.1007/JHEP05(2011)034}},
  \href{http://www.arXiv.org/abs/1103.2765}{\texttt{ arXiv:1103.2765}}.

\bibitem{top_afb_implications1}
C.~Delaunay\hrefCMSnoop {} { {et~al.}, ``{Implications of the CDF $t
  \overline{t}$ forward-backward asymmetry for hard top physics}'',} \textit{
  JHEP} \textbf{ 08} (2011) 031,
  \href{http://dx.doi.org/10.1007/JHEP08(2011)031}{\doi{10.1007/JHEP08(2011)031}},
  \href{http://www.arXiv.org/abs/1103.2297}{\texttt{ arXiv:1103.2297}}.

\bibitem{cmsdetector}
\hrefCMSnoop {} {{ CMS} Collaboration, ``The {CMS} experiment at the {CERN}
  {LHC}'',} \textit{ JINST} \textbf{ 3} (2008) S08004,
\href{http://dx.doi.org/10.1088/1748-0221/3/08/S08004}{\doi{10.1088/1748-0221/3/08/S08004}}.
%%CITATION = JINST,3,S08004;%%.

\bibitem{Deliot:2013gla}
\hrefCMSnoop {} {{ CDF/D0} Collaboration, ``{Top quark physics at the
  Tevatron}'',} \textit{ Int. J. Mod. Phys. A} \textbf{ 08} (2013) 1330013,
  \href{http://dx.doi.org/10.1142/S0217751X13300135}{\doi{10.1142/S0217751X13300135}},
\href{http://www.arXiv.org/abs/1302.3628}{\texttt{ arXiv:1302.3628}}.
%%CITATION = ARXIV:1302.3628;%%.

\bibitem{cms-allhad}
\hrefCMSnoop {} {{ CMS} Collaboration, ``{Search for anomalous $t\overline{t}$
  production in the highly-boosted all-hadronic final state}'',} \textit{ JHEP}
  \textbf{ 09} (2012) 029,
  \href{http://dx.doi.org/10.1007/JHEP09(2012)029}{\doi{10.1007/JHEP09(2012)029}},
\href{http://www.arXiv.org/abs/1204.2488}{\texttt{ arXiv:1204.2488}}.
%%CITATION = ARXIV:1204.2488;%%.

\bibitem{atlas-ljets}
\hrefCMSnoop {} {{ ATLAS} Collaboration, ``{A search for $t\overline{t}$
  resonances in the lepton plus jets final state with ATLAS using 4.7 fb$^{-1}$
  of pp collisions at $\sqrt{s} =$ 7 TeV}'',} \textit{ Phys. Rev. D} \textbf{
  88} (2013) 012004,
  \href{http://dx.doi.org/10.1103/PhysRevD.88.012004}{\doi{10.1103/PhysRevD.88.012004}},
\href{http://www.arXiv.org/abs/1305.2756}{\texttt{ arXiv:1305.2756}}.
%%CITATION = ARXIV:1305.2756;%%.

\bibitem{atlas-ljets-boosted}
\hrefCMSnoop {} {{ ATLAS} Collaboration, ``{A search for $t\overline{t}$
  resonances in lepton+jets events with highly boosted top quarks collected in
  $pp$ collisions at $\sqrt{s} = 7$ TeV with the ATLAS detector}'',} \textit{
  JHEP} \textbf{ 09} (2012) 041,
  \href{http://dx.doi.org/10.1007/JHEP09(2012)041}{\doi{10.1007/JHEP09(2012)041}},
\href{http://www.arXiv.org/abs/1207.2409}{\texttt{ arXiv:1207.2409}}.
%%CITATION = ARXIV:1207.2409;%%.

\bibitem{cms-ljets}
\hrefCMSnoop {} {{ CMS} Collaboration, ``{Search for resonant $t\bar{t}$
  production in lepton+jets events in $pp$ collisions at $\sqrt{s}=7$ TeV}'',}
  \textit{ JHEP} \textbf{ 12} (2012) 015,
  \href{http://dx.doi.org/10.1007/JHEP12(2012)015}{\doi{10.1007/JHEP12(2012)015}},
\href{http://www.arXiv.org/abs/1209.4397}{\texttt{ arXiv:1209.4397}}.
%%CITATION = ARXIV:1209.4397;%%.

\bibitem{cms-dilep}
\hrefCMSnoop {} {{ CMS} Collaboration, ``{Search for $Z^{\prime}$ resonances
  decaying to $t\bar{t}$ in dilepton+jets final states in $pp$ collisions at
  $\sqrt{s}=7$ TeV}'',} \textit{ Phys. Rev. D} \textbf{ 87} (2013) 072002,
  \href{http://dx.doi.org/10.1103/PhysRevD.87.072002}{\doi{10.1103/PhysRevD.87.072002}},
\href{http://www.arXiv.org/abs/1211.3338}{\texttt{ arXiv:1211.3338}}.
%%CITATION = ARXIV:1211.3338;%%.

\bibitem{Davoudiasl:2009cd}
\hrefCMSnoop {} {H.~Davoudiasl, S.~Gopalakrishna, E.~Ponton, and J.~Santiago,
  ``{Warped 5-dimensional models: phenomenological status and experimental
  prospects}'',} \textit{ New J. Phys.} \textbf{ 12} (2010) 075011,
  \href{http://dx.doi.org/10.1088/1367-2630/12/7/075011}{\doi{10.1088/1367-2630/12/7/075011}},
\href{http://www.arXiv.org/abs/0908.1968}{\texttt{ arXiv:0908.1968}}.
%%CITATION = ARXIV:0908.1968;%%.

\bibitem{pfalgo}
\href {http://cdsweb.cern.ch/record/1194487} {{ CMS} Collaboration,
  ``Particle--Flow Event Reconstruction in {CMS} and Performance for Jets,
  Taus, and {\MET}'',} CMS Physics Analysis Summary CMS-PAS-PFT-09-001, 2009.

\bibitem{FastJet}
\hrefCMSnoop {} {M.~Cacciari, G.~P. Salam, and G.~Soyez, ``{FastJet user
  manual}'',} \textit{ Eur. Phys. J. C} \textbf{ 72} (2012) 1896,
  \href{http://dx.doi.org/10.1140/epjc/s10052-012-1896-2}{\doi{10.1140/epjc/s10052-012-1896-2}},
\href{http://www.arXiv.org/abs/1111.6097}{\texttt{ arXiv:1111.6097}}.
%%CITATION = ARXIV:1111.6097;%%.

\bibitem{Cacciari:2008gp}
\hrefCMSnoop {} {M.~Cacciari, G.~P. Salam, and G.~Soyez, ``{The anti-$k_t$ jet
  clustering algorithm}'',} \textit{ JHEP} \textbf{ 04} (2008) 063,
  \href{http://dx.doi.org/10.1088/1126-6708/2008/04/063}{\doi{10.1088/1126-6708/2008/04/063}},
\href{http://www.arXiv.org/abs/0802.1189}{\texttt{ arXiv:0802.1189}}.
%%CITATION = 0802.1189;%%.

\bibitem{CACluster1}
\hrefCMSnoop {} {Y.~L. Dokshitzer, G.~D. Leder, S.~Moretti, and B.~R. Webber,
  ``{Better jet clustering algorithms}'',} \textit{ JHEP} \textbf{ 08} (1997)
  001,
  \href{http://dx.doi.org/10.1088/1126-6708/1997/08/001}{\doi{10.1088/1126-6708/1997/08/001}},
\href{http://www.arXiv.org/abs/hep-ph/9707323}{\texttt{ arXiv:hep-ph/9707323}}.
%%CITATION = HEP-PH/9707323;%%.

\bibitem{CACluster2}
\href {http://www.desy.de/~heramc/proceedings/wg40/} {M.~Wobisch and
  T.~Wengler, ``{Hadronization corrections to jet cross sections in deep-
  inelastic scattering}'',} (1999).
  \href{http://www.arXiv.org/abs/hep-ph/9907280}{\texttt{
  arXiv:hep-ph/9907280}}.
DESY-PROC 1999-02.
%%CITATION = HEP-PH/9907280;%%.

\bibitem{Chatrchyan:2012jua}
\hrefCMSnoop {} {{ CMS} Collaboration, ``{Identification of b-quark jets with
  the CMS experiment}'',} \textit{ JINST} \textbf{ 8} (2013) P04013,
  \href{http://dx.doi.org/10.1088/1748-0221/8/04/P04013}{\doi{10.1088/1748-0221/8/04/P04013}},
\href{http://www.arXiv.org/abs/1211.4462}{\texttt{ arXiv:1211.4462}}.
%%CITATION = ARXIV:1211.4462;%%.

\bibitem{TopTagAlgo}
\hrefCMSnoop {} {D.~E. Kaplan, K.~Rehermann, M.~D. Schwartz, and B.~Tweedie,
  ``{Top Tagging: A method for identifying boosted hadronically decaying top
  quarks}'',} \textit{ Phys. Rev. Lett.} \textbf{ 101} (2008) 142001,
  \href{http://dx.doi.org/10.1103/PhysRevLett.101.142001}{\doi{10.1103/PhysRevLett.101.142001}},
\href{http://www.arXiv.org/abs/0806.0848}{\texttt{ arXiv:0806.0848}}.
%%CITATION = 0806.0848;%%.

\bibitem{powheg}
\hrefCMSnoop {} {S.~Alioli, P.~Nason, C.~Oleari, and E.~Re, ``{A general
  framework for implementing NLO calculations in shower Monte Carlo programs:
  the POWHEG BOX}'',} \textit{ JHEP} \textbf{ 06} (2010) 043,
  \href{http://dx.doi.org/10.1007/JHEP06(2010)043}{\doi{10.1007/JHEP06(2010)043}},
\href{http://www.arXiv.org/abs/1002.2581}{\texttt{ arXiv:1002.2581}}.
%%CITATION = ARXIV:1002.2581;%%.

\bibitem{pythia}
\hrefCMSnoop {} {T.~Sj{\"o}strand, S.~Mrenna, and P.~Z. Skands, ``{PYTHIA 6.4
  physics and manual}'',} \textit{ JHEP} \textbf{ 05} (2006) 026,
  \href{http://dx.doi.org/10.1088/1126-6708/2006/05/026}{\doi{10.1088/1126-6708/2006/05/026}},
\href{http://www.arXiv.org/abs/hep-ph/0603175}{\texttt{ arXiv:hep-ph/0603175}}.
%%CITATION = HEP-PH/0603175;%%.

\bibitem{Chatrchyan:2011id}
\hrefCMSnoop {} {{ CMS} Collaboration, ``{Measurement of the underlying event
  activity at the LHC with $\sqrt{s}= 7$ TeV and comparison with $\sqrt{s} =
  0.9$ TeV}'',} \textit{ JHEP} \textbf{ 09} (2011) 109,
  \href{http://dx.doi.org/10.1007/JHEP09(2011)109}{\doi{10.1007/JHEP09(2011)109}},
\href{http://www.arXiv.org/abs/1107.0330}{\texttt{ arXiv:1107.0330}}.
%%CITATION = ARXIV:1107.0330;%%.

\bibitem{madgraph}
J.~Alwall\hrefCMSnoop {} { {et~al.}, ``{MadGraph 5: Going beyond}'',} \textit{
  JHEP} \textbf{ 06} (2011) 128,
  \href{http://dx.doi.org/10.1007/JHEP06(2011)128}{\doi{10.1007/JHEP06(2011)128}},
  \href{http://www.arXiv.org/abs/1106.0522}{\texttt{ arXiv:1106.0522}}.

\bibitem{pythia8}
\hrefCMSnoop {} {T.~Sj{\"o}strand, S.~Mrenna, and P.~Z. Skands, ``{A brief
  introduction to PYTHIA 8.1}'',} \textit{ Comput. Phys. Commun.} \textbf{ 178}
  (2008) 852,
  \href{http://dx.doi.org/10.1016/j.cpc.2008.01.036}{\doi{10.1016/j.cpc.2008.01.036}},
\href{http://www.arXiv.org/abs/0710.3820}{\texttt{ arXiv:0710.3820}}.
%%CITATION = ARXIV:0710.3820;%%.

\bibitem{Nadolsky:2008zw}
P.~M. Nadolsky\hrefCMSnoop {} { {et~al.}, ``Implications of {CTEQ} global
  analysis for collider observables'',} \textit{ Phys. Rev. D} \textbf{ 78}
  (2008) 013004,
  \href{http://dx.doi.org/10.1103/PhysRevD.78.013004}{\doi{10.1103/PhysRevD.78.013004}},
\href{http://www.arXiv.org/abs/0802.0007}{\texttt{ arXiv:0802.0007}}.
%%CITATION = 0802.0007;%%.

\bibitem{CT10}
H.-L. Lai\hrefCMSnoop {} { {et~al.}, ``New parton distributions for collider
  physics'',} \textit{ Phys. Rev. D} \textbf{ 82} (2010) 074024,
  \href{http://dx.doi.org/10.1103/PhysRevD.82.074024}{\doi{10.1103/PhysRevD.82.074024}},
\href{http://www.arXiv.org/abs/1007.2241}{\texttt{ arXiv:1007.2241}}.
%%CITATION = ARXIV:1007.2241;%%.

\bibitem{LumiUncertainty}
\href {http://cds.cern.ch/record/1598864} {{ CMS} Collaboration, ``{CMS}
  luminosity based on pixel cluster counting - {S}ummer 2013 update'',} CMS
  Physics Analysis Summary CMS-PAS-LUM-13-001, 2013.

\bibitem{PhysRevD.84.092004}
\hrefCMSnoop {} {{ CMS} Collaboration, ``{Measurement of the $t \bar{t}$
  production cross section in $pp$ collisions at 7 TeV in lepton + jets events
  using $b$-quark jet identification}'',} \textit{ Phys. Rev. D} \textbf{ 84}
  (2011) 092004,
  \href{http://dx.doi.org/10.1103/PhysRevD.84.092004}{\doi{10.1103/PhysRevD.84.092004}},
\href{http://www.arXiv.org/abs/1108.3773}{\texttt{ arXiv:1108.3773}}.
%%CITATION = ARXIV:1108.3773;%%.

\bibitem{Pumplin:2001ct}
J.~Pumplin\hrefCMSnoop {} { {et~al.}, ``{Uncertainties of predictions from
  parton distribution functions. 2. The Hessian method}'',} \textit{ Phys. Rev.
  D} \textbf{ 65} (2001) 014013,
  \href{http://dx.doi.org/10.1103/PhysRevD.65.014013}{\doi{10.1103/PhysRevD.65.014013}},
\href{http://www.arXiv.org/abs/hep-ph/0101032}{\texttt{ arXiv:hep-ph/0101032}}.
%%CITATION = HEP-PH/0101032;%%.

\bibitem{Gao:2010bb}
J.~Gao\hrefCMSnoop {} { {et~al.}, ``{Next-to-leading order QCD corrections to
  the heavy resonance production and decay into top quark pair at the LHC}'',}
  \textit{ Phys. Rev. D} \textbf{ 82} (2010) 014020,
  \href{http://dx.doi.org/10.1103/PhysRevD.82.014020}{\doi{10.1103/PhysRevD.82.014020}},
\href{http://www.arXiv.org/abs/1004.0876}{\texttt{ arXiv:1004.0876}}.
%%CITATION = ARXIV:1004.0876;%%.

\end{thebibliography}\endgroup
\ifthenelse{\boolean{cms@external}}{}{
\clearpage
\appendix
\newlength\cmsFigWidthSupp
\ifthenelse{\boolean{cms@external}}{\setlength\cmsFigWidthSupp{0.95\columnwidth}}{\setlength\cmsFigWidthSupp{0.48\columnwidth}}

\ifthenelse{\boolean{cms@external}}{}{\section{Supplemental Material \label{app:suppMat}}}

\begin{figure}[t]
\includegraphics[width=\cmsFigWidthSupp]{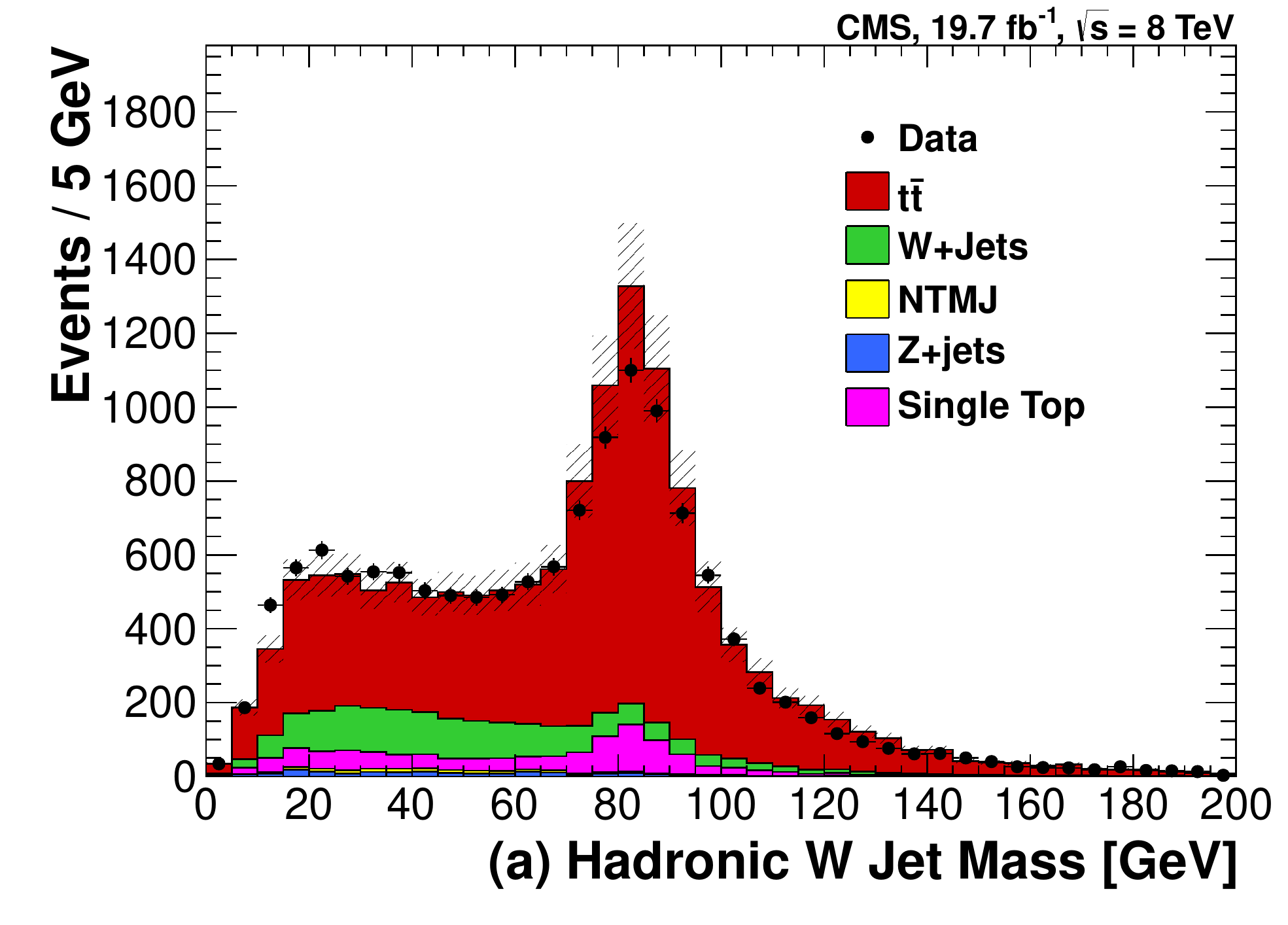}
\includegraphics[width=\cmsFigWidthSupp]{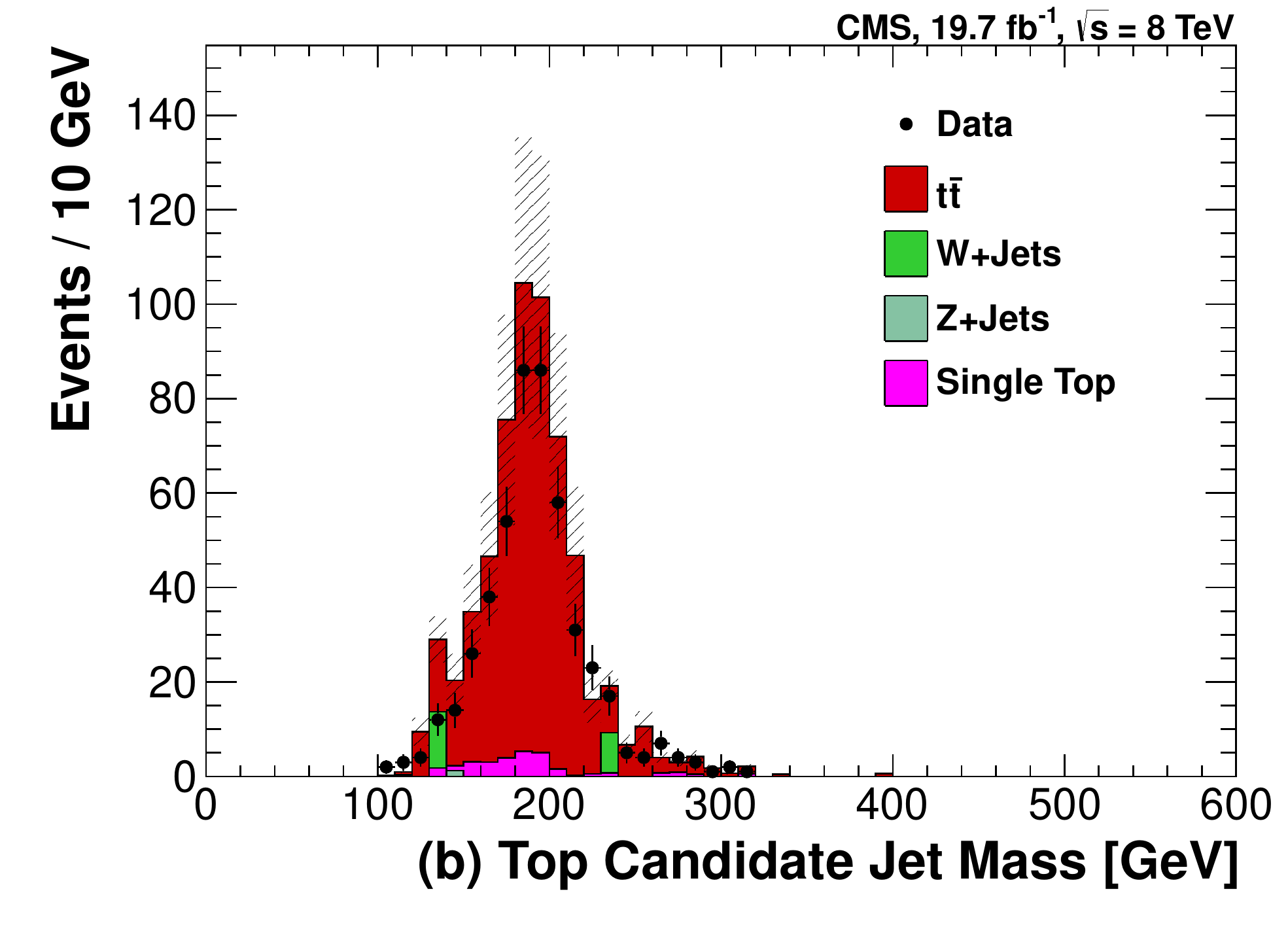}
\caption{Jet mass distribution for fully-merged W decay products (a) and fully-merged top quark candidate jets (b), in the muon + jets selection.  The shaded band corresponds to the total SM background uncertainty.
\label{fig:mass}}
\end{figure}
\ifthenelse{\boolean{cms@external}}
{\vspace{-0.3in}
}
{
}
\noindent
We provide additional plots that illustrate the top-tagging procedure used in the analysis presented in this Letter.
Figure \ref{fig:mass} shows the distribution of single-jet masses in a selection optimized to identify partially-merged top quark decays. 
In this topology, the W boson decay products will be merged into a
single jet, but the $b$ quark will escape. 
\ifthenelse{\boolean{cms@external}}{\pagebreak}{}
\begin{figure}[t!p]
\includegraphics[width=0.48\columnwidth]{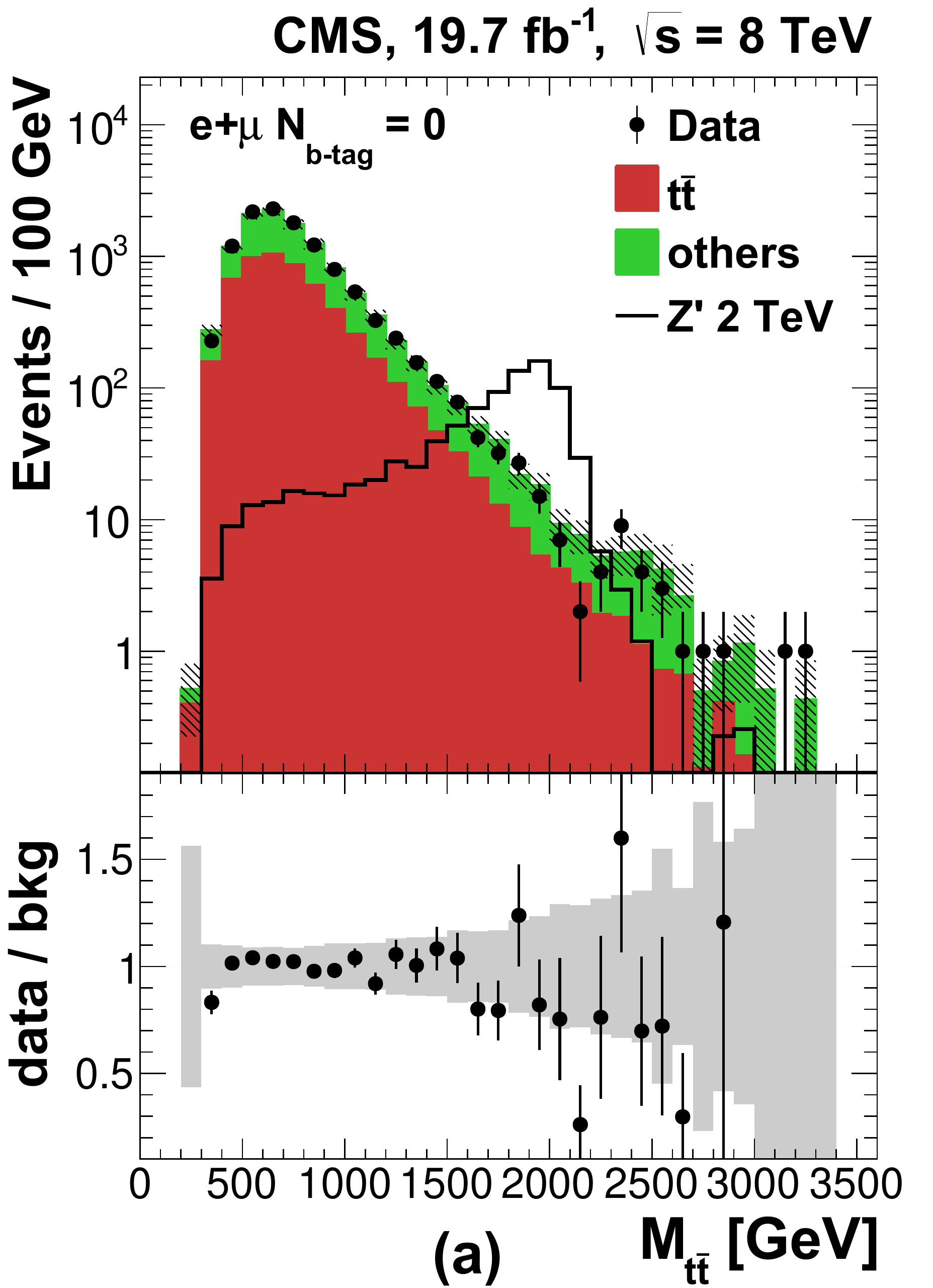}
\includegraphics[width=0.48\columnwidth]{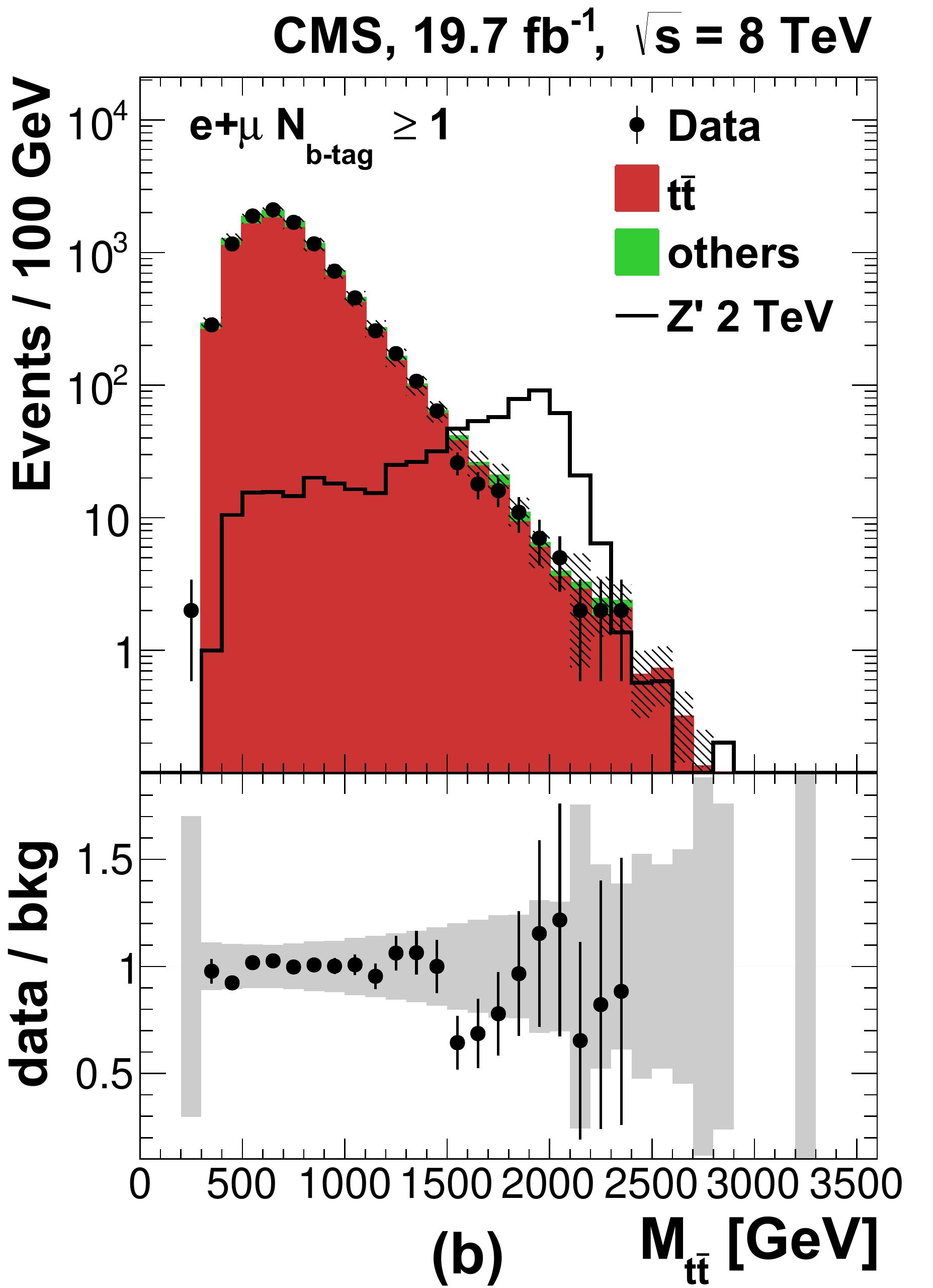}\\
\includegraphics[width=0.48\columnwidth]{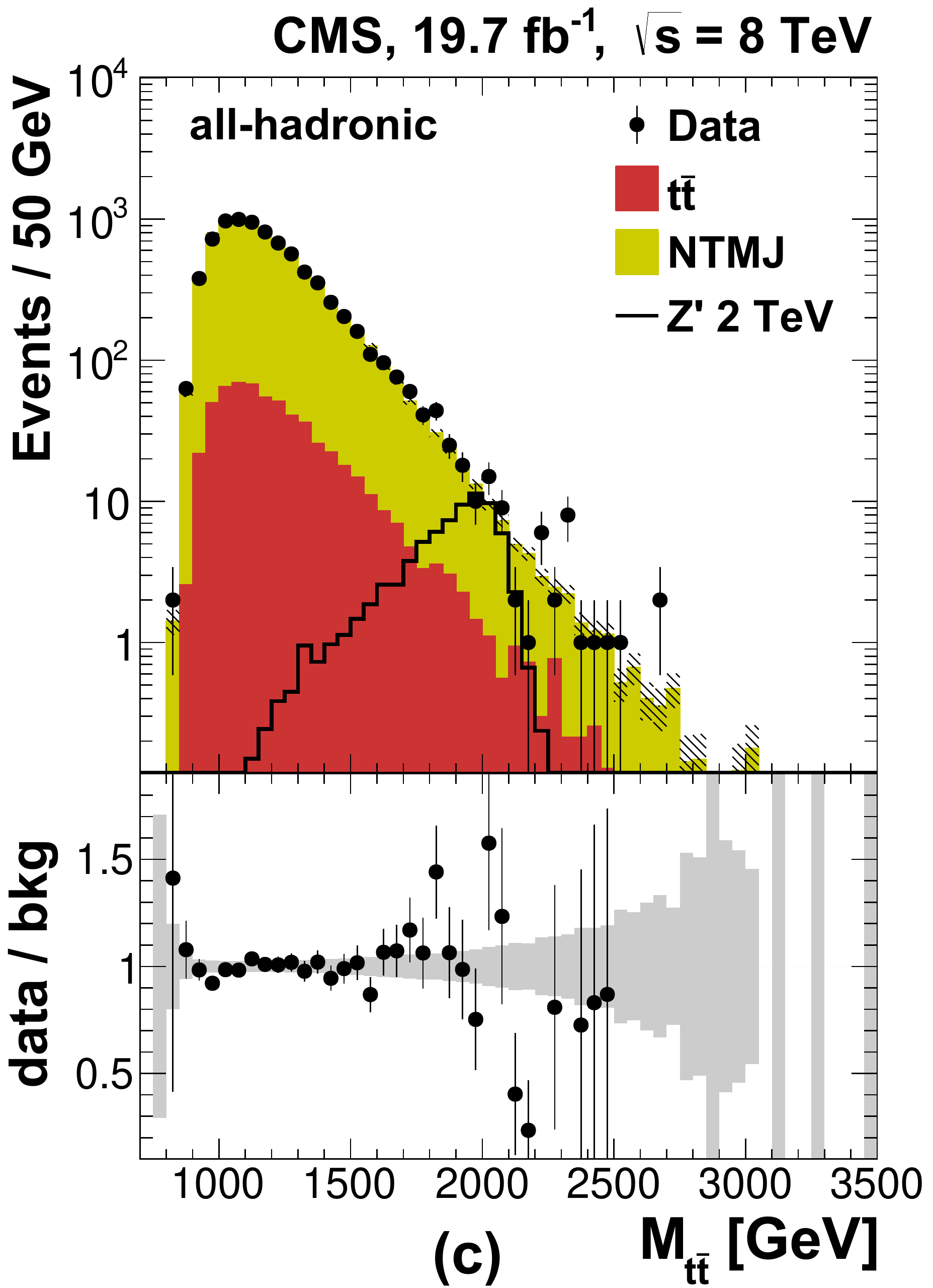}
\includegraphics[width=0.48\columnwidth]{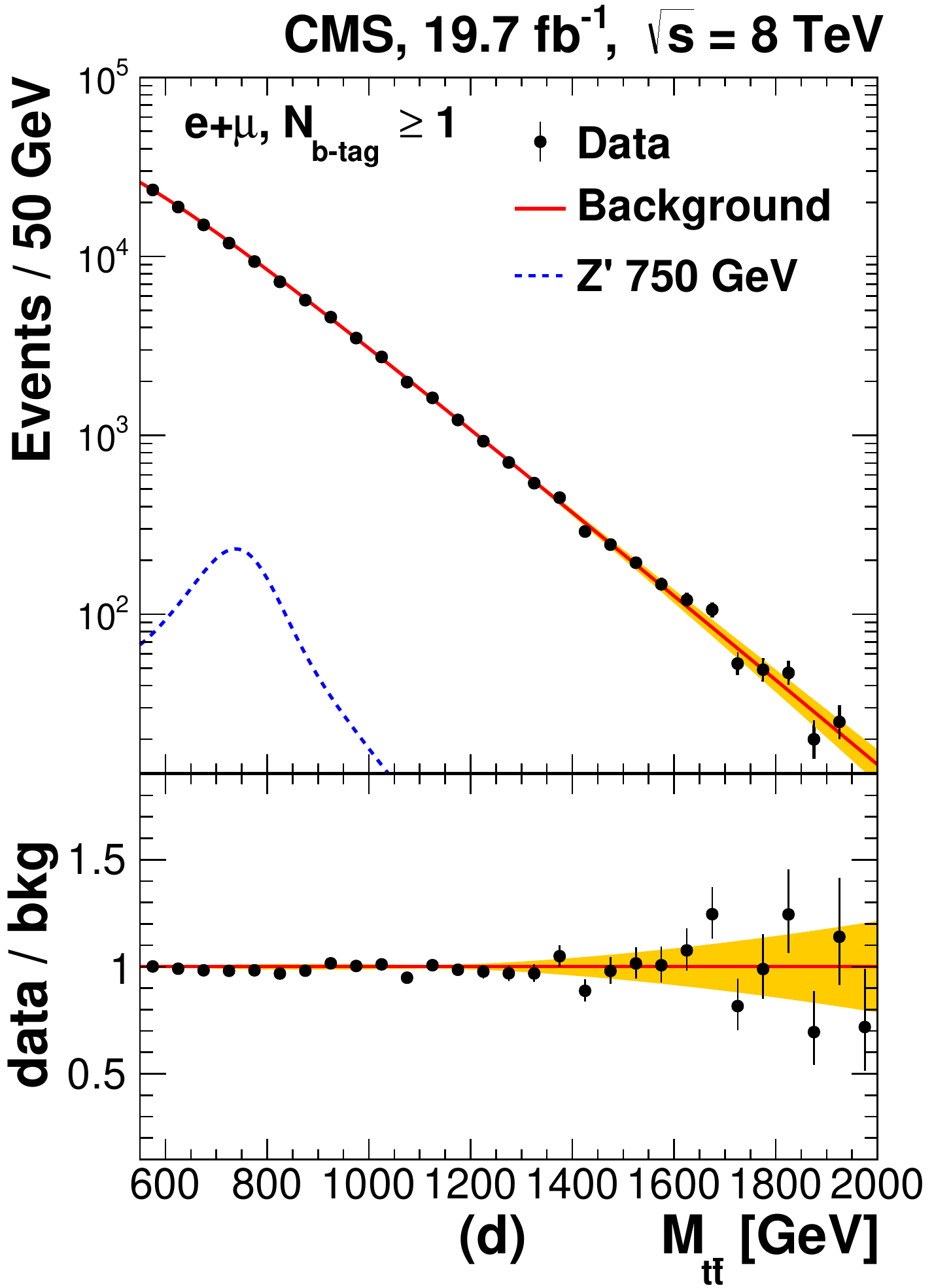}
\ifthenelse{\boolean{cms@external}}
{\caption{Plots from Fig.~1 from the Letter, including plots of the ratio of data and total expected background for each analysis channel.}
}
{\caption{Plots from Fig.~\ref{fig:mttbar}, including plots of the ratio of data and total expected background for each analysis channel.}
}
\label{fig:mttbar_ratios}
\end{figure}
\noindent
The figure also shows the jet mass distribution for fully-merged boosted top quarks. 
The reconstruction of these kinematic observables, in a sample enriched in $t\bar{t}$ events in data, serves to validate the top-tagging algorithm. Further details are given in 
\ifthenelse{\boolean{cms@external}}{Ref.~20}{Ref.~\cite{cms-allhad}} of the Letter.
We also repeat the plots from Fig.~1 of the main text here in Fig.~\ref{fig:mttbar_ratios}, including ratios to better show the agreement between data and expectation.

}

\cleardoublepage \section{The CMS Collaboration \label{app:collab}}\begin{sloppypar}\hyphenpenalty=5000\widowpenalty=500\clubpenalty=5000\textbf{Yerevan Physics Institute,  Yerevan,  Armenia}\\*[0pt]
S.~Chatrchyan, V.~Khachatryan, A.M.~Sirunyan, A.~Tumasyan
\vskip\cmsinstskip
\textbf{Institut f\"{u}r Hochenergiephysik der OeAW,  Wien,  Austria}\\*[0pt]
W.~Adam, T.~Bergauer, M.~Dragicevic, J.~Er\"{o}, C.~Fabjan\cmsAuthorMark{1}, M.~Friedl, R.~Fr\"{u}hwirth\cmsAuthorMark{1}, V.M.~Ghete, N.~H\"{o}rmann, J.~Hrubec, M.~Jeitler\cmsAuthorMark{1}, W.~Kiesenhofer, V.~Kn\"{u}nz, M.~Krammer\cmsAuthorMark{1}, I.~Kr\"{a}tschmer, D.~Liko, I.~Mikulec, D.~Rabady\cmsAuthorMark{2}, B.~Rahbaran, C.~Rohringer, H.~Rohringer, R.~Sch\"{o}fbeck, J.~Strauss, A.~Taurok, W.~Treberer-Treberspurg, W.~Waltenberger, C.-E.~Wulz\cmsAuthorMark{1}
% ...
\vskip\cmsinstskip
% ...
1:~~Also at Vienna University of Technology, Vienna, Austria\\
2:~~Also at CERN, European Organization for Nuclear Research, Geneva, Switzerland\\
% ...

\end{sloppypar}
\end{document}